\documentclass[preprints,article,accept,moreauthors,pdftex]{Definitions/mdpi}
\newcommand{\EX}{\mathrm{E}}
\firstpage{1} 

\pubvolume{2}
\issuenum{2}
\articlenumber{12}
\pubyear{2019}
\copyrightyear{2019}
\history{Received: 8 March 2019; Accepted: 6 May 2019; Published: 9 May 2019}

\Title{Experimental and signal processing techniques for fault diagnosis on a small horizontal-axis wind turbine generator}

\conference{International Conference on Noise and Vibration Engineering and USD 2018, 17-19 September 2018, Leuven (Belgium)}

\Author{Francesco Natili$^{1,\ddagger}$, Francesco Castellani $^{1,\ddagger}$, Davide Astolfi $^{1,\ddagger}$*, Matteo Becchetti $^{1,\ddagger}$}

\AuthorNames{Francesco Natili, Francesco Castellani, Davide Astolfi, Matteo Becchetti}

\address{%
$^{1}$ \quad  University of Perugia - Department of Engineering, Via G. Duranti 93 - 06125 Perugia (Italy); francesco.natili@yahoo.it; francesco.castellani@unipg.it; davide.astolfi@unipg.it; matteo.becchetti@unipg.it}

\corres{Correspondence: davide.astolfi@unipg.it; Tel.: +39 075 585 3709}

\secondnote{These authors contributed equally to this work.}

\abstract{Small HAWT is a technology characterized by non-trivial critical points, basically because it is targeted for domestic use and therefore cheap manufacturing and control must conjugate with good efficiency under possibly complex flow conditions (especially in urban environment). Therefore, dynamical control optimization and noise and vibration mitigation are pressing issues for this kind of technology. Despite it is peculiar of small HAWTs that the generator constitutes a non-negligible fraction of the total mass and therefore the electromechanical coupling is relevant, condition monitoring of small HAWT generators is an overlooked topic. The present work is a test case study of damage diagnosis on a permanent magnet generator of a HAWT having 3 kW of maximum power and 2 meters of rotor diameter. The experimental analysis is conducted through wind tunnel tests and on a generator test rig where a damaged and an undamaged generators have been driven at different rotational speeds. Vibration measurements are collected in the wind tunnel through radial accelerometers near the rear bearing of the shaft and in the test rig through uni-axial accelerometers (fixed in radial positions, in order to be aligned with front and rear bearings). The test rig data result being particularly useful for studying the low-frequency tail of the vibration spectrum, where the characteristic frequencies of the bearing are located. The experimental data are analyzed in the time and frequency domain for feature extraction: a fault in the cage of the bearing supporting the generator is diagnosed using in particular the spectral coherence analysis.}


\begin{document}

\section{Introduction}
\label{S1}
A reliable and efficient use of small HAWT technology \cite{tummala2016review} poses several challenges, because on one hand the market requests low price for domestic investment (and this means cheap manufacturing and control) and, on the other hand, the flow conditions to which these devices are commonly subjected (especially in urban environment \cite{sang2017experimental, tabrizi2014performance, lubitz2014impact, pagnini2015experimental}) can be particularly complex. 

In order to guarantee acceptable values of the power factor $C_p$, HAWT typically have rotational speed up to several hundreds rpm. This implies that noise and vibration control \cite{joosse2002acoustic, mollasalehi2013contribution, lee2014numerical} is a particularly pressing issue. Activities at this aim should evidently be performed in the prototyping phase, because small HAWT typically are not equipped with condition monitoring systems.

As observed, for example, in \cite{castellani2018experimental}, the mechanical behavior of small HAWTs is strongly affected by the importance of the electromechanical couplings, due to fact that the generator constitutes a remarkable fraction of the total mass of the device. Therefore, the generator can transmit to the small HAWT structure powerful vibrations, especially at high frequency. Despite this matter of fact, the literature about condition monitoring and fault diagnosis on small HAWT generators is remarkably missing. 

Mollasalehi, Wood and Sun have devoted particular attention to the study of wind turbine generator bearing damages. Evidently, due to the energy scale involved, the main field of application are MW-scale wind turbines. In \cite{cao2018prediction}, for example, a data-driven method is proposed for estimating generator bearings remaining useful life, combining the interval whitenization method with a Gaussian process algorithm. In \cite{mollasalehi2017indicative}, a method is proposed for indicative fault diagnosis on wind turbine generator bearings: it is based on sound and vibration analysis at the bottom of the tower and the key point is extracting information about what happens at the top tower through measurements at the bottom (that are much easier and less expensive by a practical point view). In \cite{Mollasalehi201493}, demodulation vibration analysis of the generator bearing of a large wind turbine is performed using the highest energy band calculated by wavelet packet transform. It is argued that the results from vibration analysis are consistent with a damage on the outer race of the bearing.

As regards condition monitoring of generator bearings of small HAWT, the only contribution to the topic, to the best of the author's knowledge, is \cite{cai2016condition} and also in that work it is noticed that the topic is very overlooked. In \cite{cai2016condition}, fault diagnosis on a 500 W small HAWT permanent magnet generator is addressed. The measurements are conducted in an ad-hoc test rig, allowing to run the generator at different shaft speeds. Two defects (ball bearing outer race defect and rotor imbalance) are detected using several system processing techniques (continuous wavelet transform, short-time Fourier transform, order analysis, wavelet power spectrum).

On these grounds, this work aims at furnishing a contribution to the topic of condition monitoring of small HAWT generators through the analysis of experimental data. The qualifying point of this study is that measurements are collected using a test rig (similarly to \cite{cai2016condition}) on the wind turbine sub-component of interest and in the wind tunnel, running the whole wind energy conversion device (a three-bladed HAWT having 3 kW of maximum power and 2 meters of rotor diameter \cite{scappatici2016optimizing}). The objective of this research is therefore basically interpreting the wind tunnel measurements in light of the test rig measurements. This is interesting because it would be quicker and cost-saving to have at hand condition monitoring strategies for small HAWTs that are based on wind tunnel tests on the whole device. Nevertheless, the vibration spectra of these kind of devices are particularly complex and difficult to interpret. This is basically due to the devices size: because of the rotor size, small HAWTs have a severely variable operating range (from dozens to several hundreds of rotor revolutions per minute); because of the device total mass, the generator is a remarkable fraction and therefore the electromechanical coupling constitutes a relevant phenomenon (as discussed, for example, in \cite{castellani2018experimental}.

The general lesson from this case study is that, despite the simplicity and the cheap manufacturing of a small HAWT, the condition monitoring for this kind of systems requires sophisticated measurements and signal processing techniques. Therefore, this work is organized as a test case discussion: it can be read as a step by step procedure for the detection of the generator malfunctioning and for identifying the possible causes. The decisive analysis for diagnosis of the anomaly has been possible using the test rig measurements, because they are more appropriate for the study of the low-frequency spectrum where the characteristic frequencies of the bearings are located. In particular, the spectral coherence analysis has resulted to be a powerful technique because it clearly highlights the frequency content of cyclostationary signals \cite{antoni2007cyclic} and the contributions to spectrum can be interpreted in relation to the theoretical characteristic frequencies of the bearing. Finally, therefore, a damage at the cage of the bearing of the generator has been identified.

The structure of the manuscript is therefore the following: in section \ref{S2}, the facilities (HAWT, wind tunnel, test rig) are briefly described; in section \ref{S3}, the damage diagnosis is conducted on the grounds of theoretical principles and signal processing techniques; the results are shortly discussed in Section \ref{disc}; finally, conclusions and further directions are drawn in the final section \ref{S4}.

\section{The test case and the facilities}
\label{S2}

\begin{figure}[H]
  \begin{center}
    \includegraphics[width=0.4\textwidth]{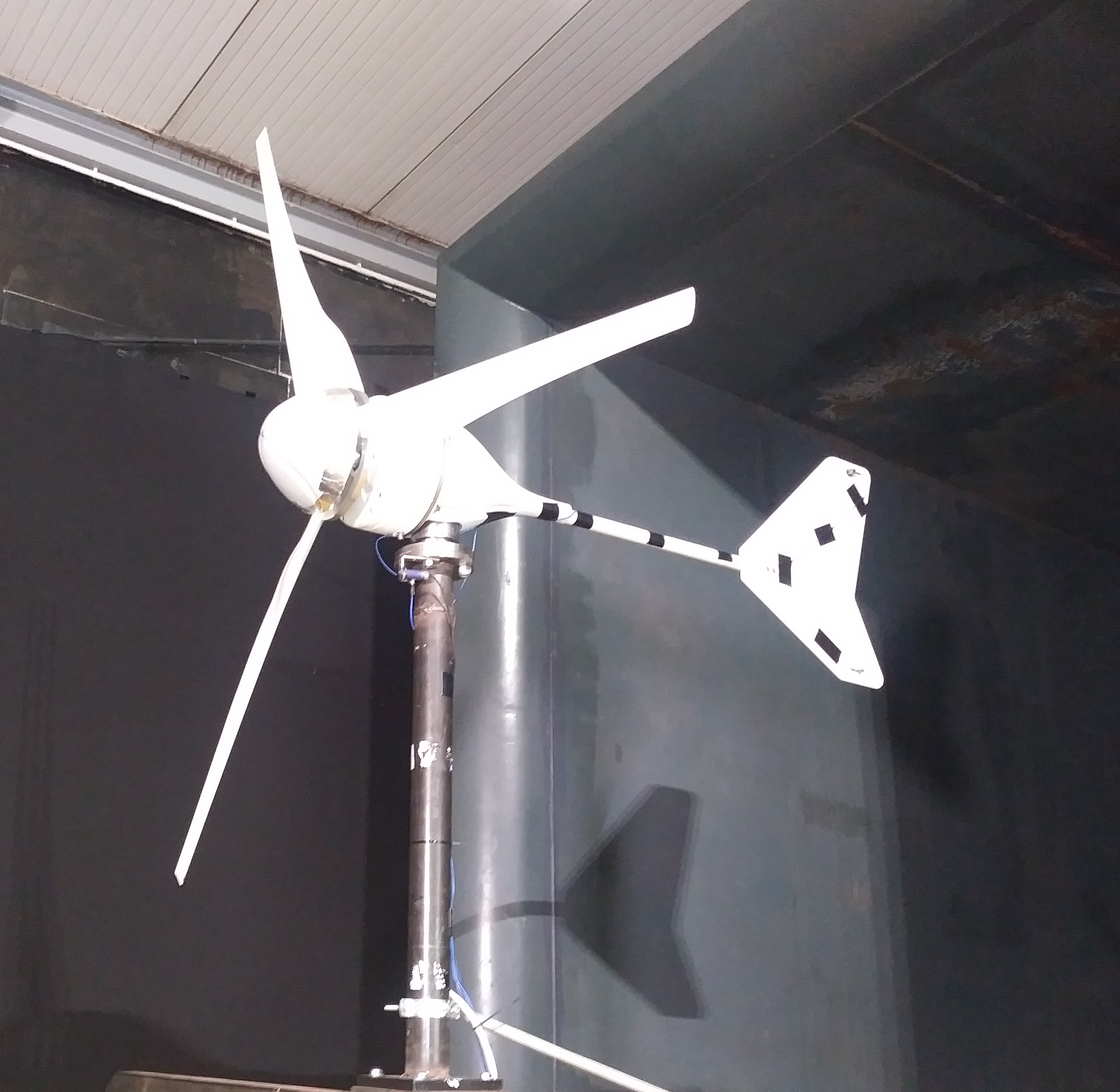}
    \caption{The test case HAWT in the wind tunnel.}
     \label{tur}
  \end{center}
\end{figure}

The test case HAWT  is three-bladed, has 2 meters of rotor diameter and the maximum producible power is 3 kW. The overall nacelle mass is 40 kg. The blades are in polymer reinforced with glass fibers and have fixed pitch angle; the minimum chord of the profile is 5 cm and the maximum is 15 cm; the minimum angle of attack for the profile is $1.7^\circ$ and the maximum is $32^\circ$.  The hub height is 1.2 meters in the wind tunnel configuration (Figure \ref{tur}). This prototype has been object of several studies about its design \cite{scappatici2016optimizing} and about its dynamic mechanical behavior \cite{castellani2017dynamic, castellani2018experimental, castellani2018experimentalmachines, castellani2019yawing}.


\begin{figure}[H]
  \begin{center}
    \includegraphics[width=0.6\textwidth]{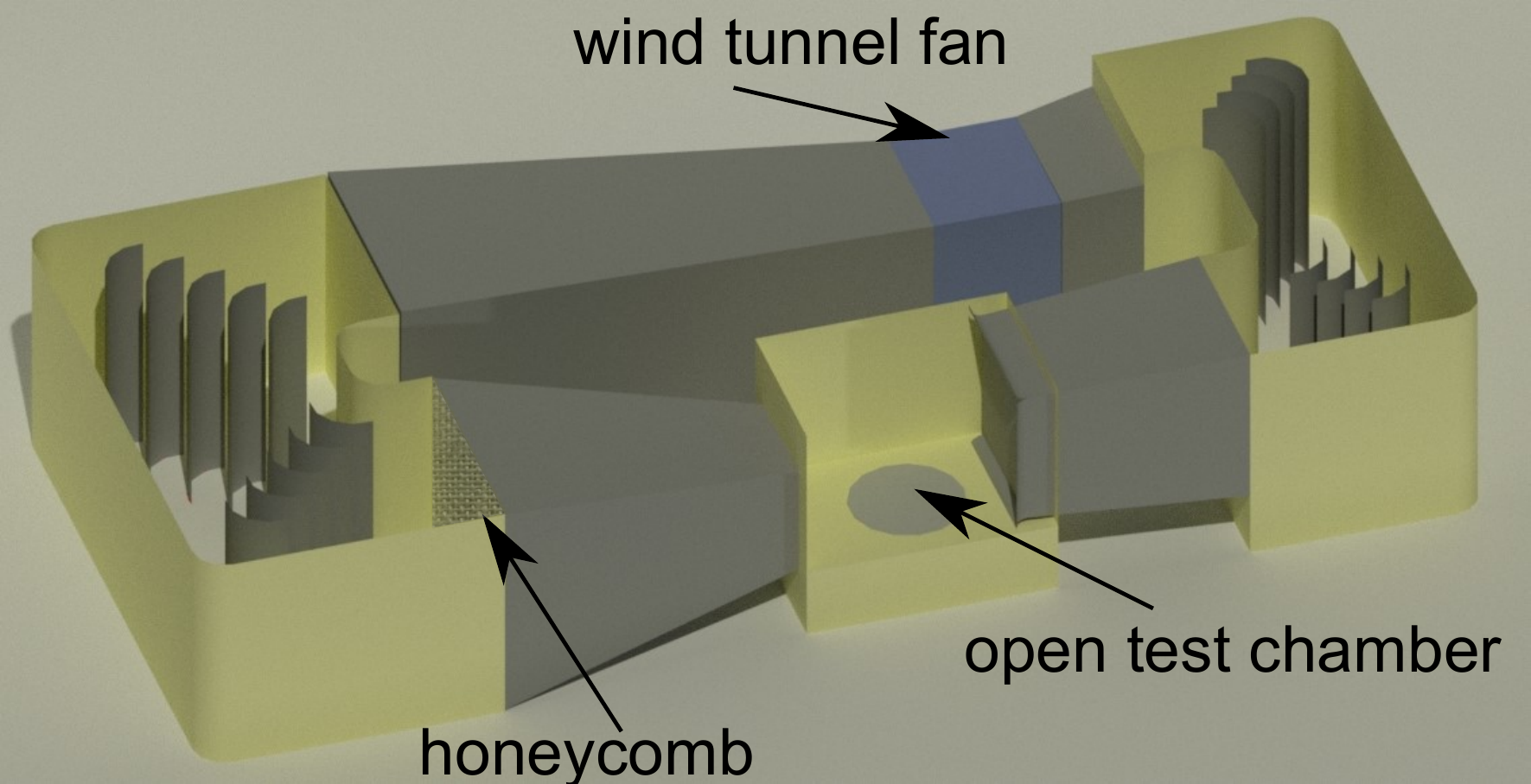}
    \caption{A scheme of the wind tunnel.}
     \label{tunnel}
  \end{center}
\end{figure}

The wind tunnel facility of the University of Perugia (www.windtunnel.unipg.it) has an open test chamber section of $2.2 \times 2.2$ meters and a recovery section of $2.7 \times 2.7$ meters. The air can be accelerated up to a maximum speed of 45 m/s using a fan driven by a 375 kW electric motor in a closed loop circuit. The level of turbulence is quite low ($<0.4\%$). The wind speed is measured by a Pitot tube and a cup anemometer placed at the inlet section. In Figure \ref{tunnel}, a scheme of the wind tunnel is reported.

During the wind tunnel tests, vibration measurements are collected through a radial accelerometer near the rear bearing of the shaft. The sampling frequency is 5 kHz. The operational conditions (basically, the rotational speed of the rotor) is contextually measured through a tachometer. 

\begin{figure}[H]
  \begin{center}
    \includegraphics[width=0.8\textwidth]{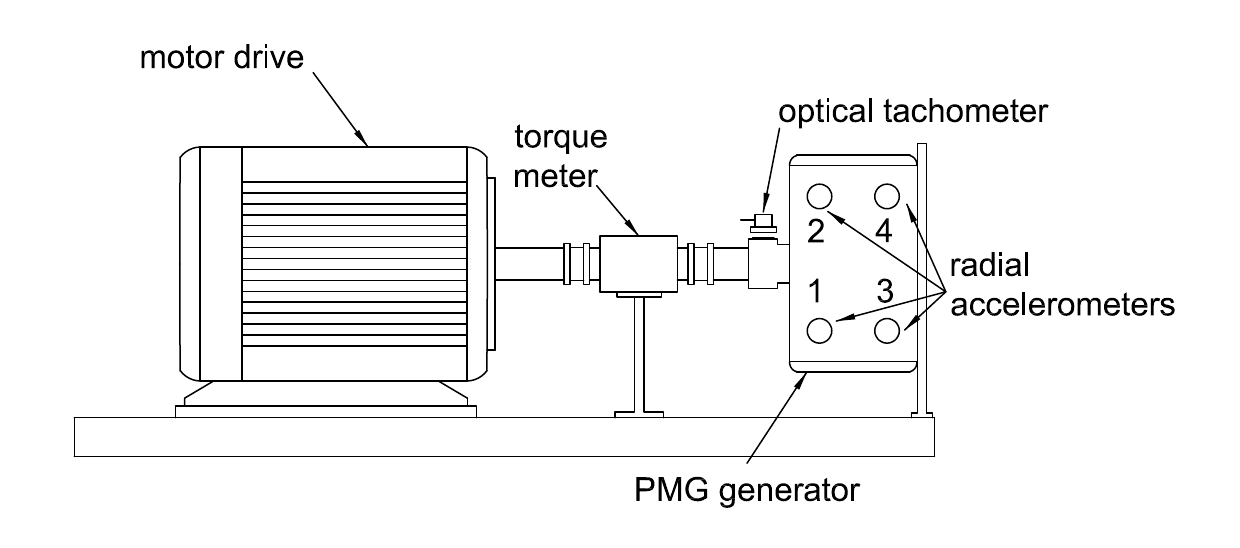}
    \caption{The test rig for the generators}
     \label{rig}
  \end{center}
\end{figure}

The test rig is displayed in Figure \ref{rig}: in the figure, the position of the employed accelerometers is indicated. On the generator, uni-axial accelerometers have been fixed in radial positions, in order to be aligned with front and rear bearings. A torque meter was installed on the shaft from the motor to the generator, while rpm were measured thanks to an optical tachometer. Also electrical parameters (voltage and current) were monitored on the brake circuit on the DC resistive load, in order to estimate the power output from the generator. The sampling frequency is 5 kHz.

Two 1.8 kW PMGs have been tested, collecting mechanical and electric data during wind tunnel ramp and steady test and driving the generators on the test rig at several shaft speeds. The two electric generators are exactly the same model, but one of them was affected by anomalous vibrations and non-optimal performances, as discussed in section \ref{S3}.

\section{The fault diagnosis}
\label{S3}

The non-optimal performances of the damaged generator have been detected as follows. In the test rig, the driving torque is provided by an electric motor, applying a resistive load on the generator
circuit side. Using the measurements from the torque
meter and the optical tachometer, it is possible to calculate the mechanical power incoming in the generator:
\begin{equation}
P_{MEC}=C_m \omega=\frac{2 \pi N C_m}{60},
\end{equation}
where $N$ is the number of revolutions per minute, $C_m$ is the torque and $\omega$ is the angular speed. As regards the electric side of the test rig, the three-phases power is transformed into direct-current using a rectifier; the voltage and current intensity are measured in order to calculate the electric power produced by the generator as 
\begin{equation}
P_{el}=V i.
\end{equation}
The efficiency of the generator can therefore be estimated as
\begin{equation}
\eta=\frac{P_{el}}{P_{mec}}=\frac{60 V i}{2 \pi N C_m}
\end{equation}
and the results are collected in Table \ref{tab1} for some representative values of rpm. It arises that the damaged generator has a lower efficiency with respect to the undamaged one. From Table \ref{tab1}, it also arises that the test rig measurements are fundamental to detect an underperformance of this order of magnitude: during the operation of the HAWT, it would be indistinguishable with respect to performance fluctuations caused by the wind flow and its interplay with the wind turbine control system.

\begin{table}[H]
\begin{center}
\caption{Mean efficiency $\eta$ at different rpm for undamaged and damaged generator.}
\label{tab1}
\begin{tabular}{c|cc}
\hline
rpm  & healthy (\%) & damaged (\%) \\
\hline
300      &    $96.55$          &    $95.68$          \\
400  &  $97.21$   &    $96.98$        \\
500  &    $97.80$        &  $96.67$   \\
\hline
    
\end{tabular}

\end{center}
\end{table}

\begin{figure}[H]
  \begin{center}
    \includegraphics[width=0.6\textwidth]{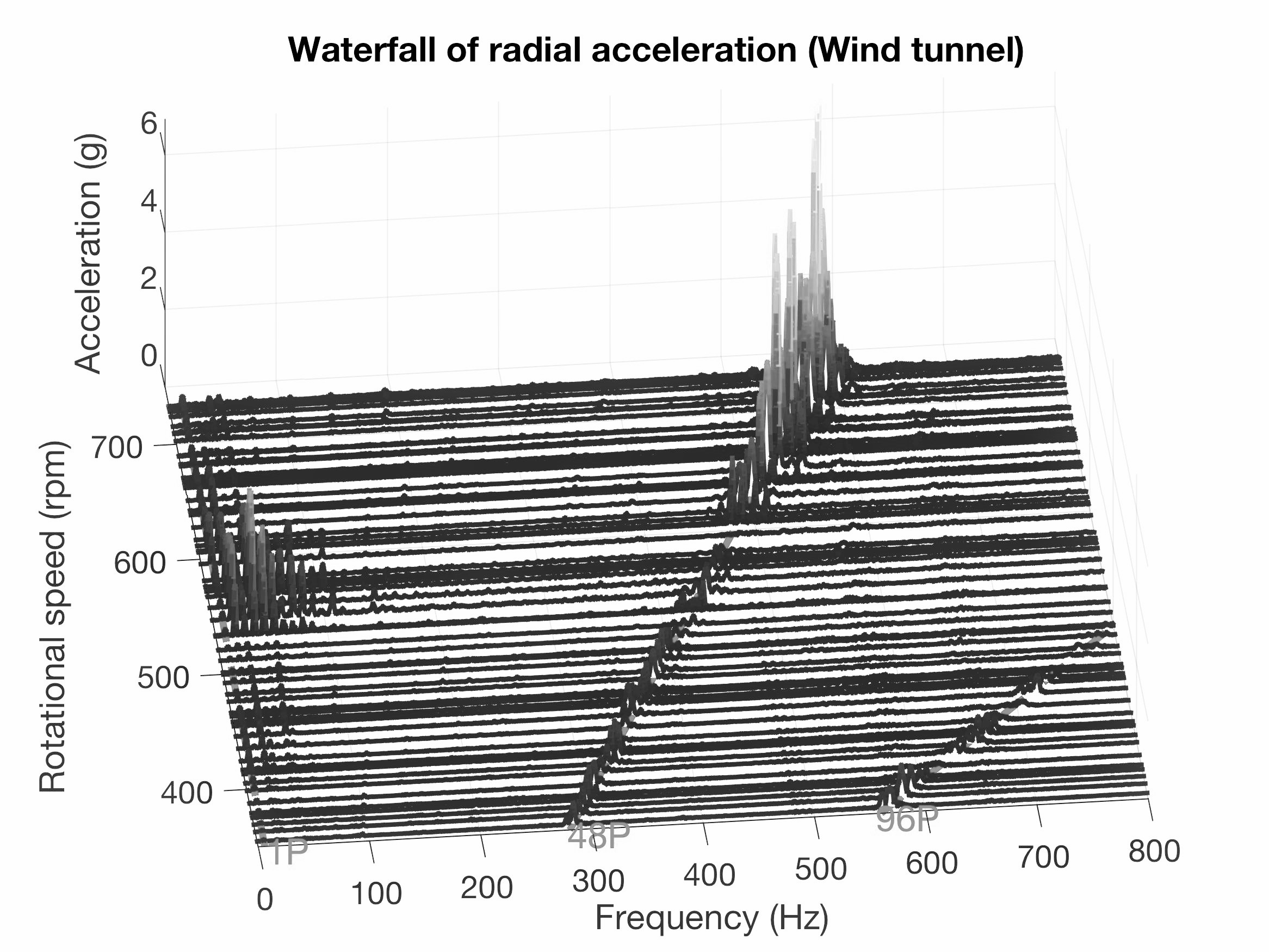}
    \caption{The waterfall plot of radial vibrations: damaged generator.}
     \label{wat1}
  \end{center}
\end{figure}

\begin{figure}[H]
  \begin{center}
    \includegraphics[width=0.6\textwidth]{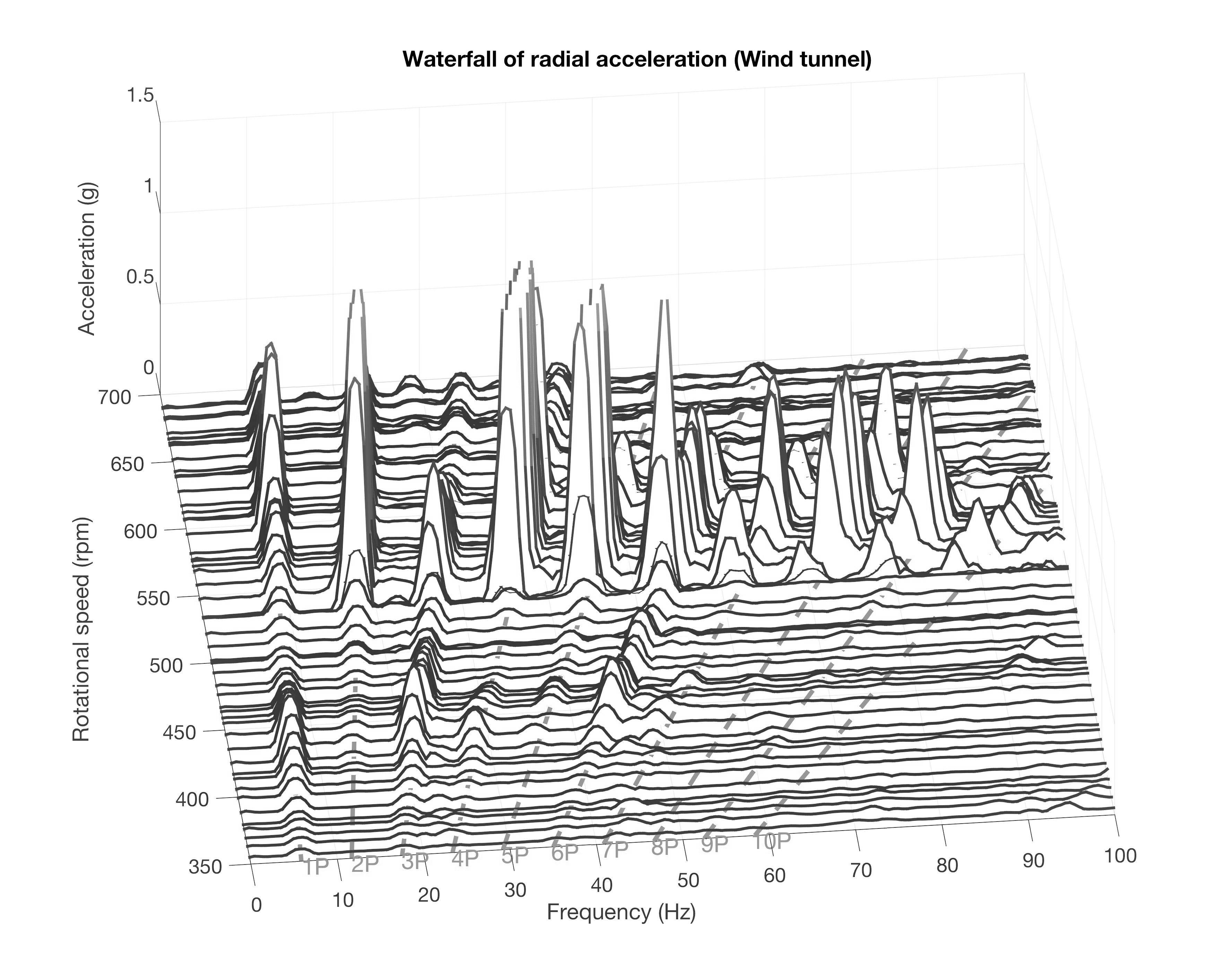}
    \caption{The waterfall plot of low frequency radial vibrations: damaged generator.}
     \label{wat2}
  \end{center}
\end{figure}

\begin{figure}[H]
  \begin{center}
    \includegraphics[width=0.6\textwidth]{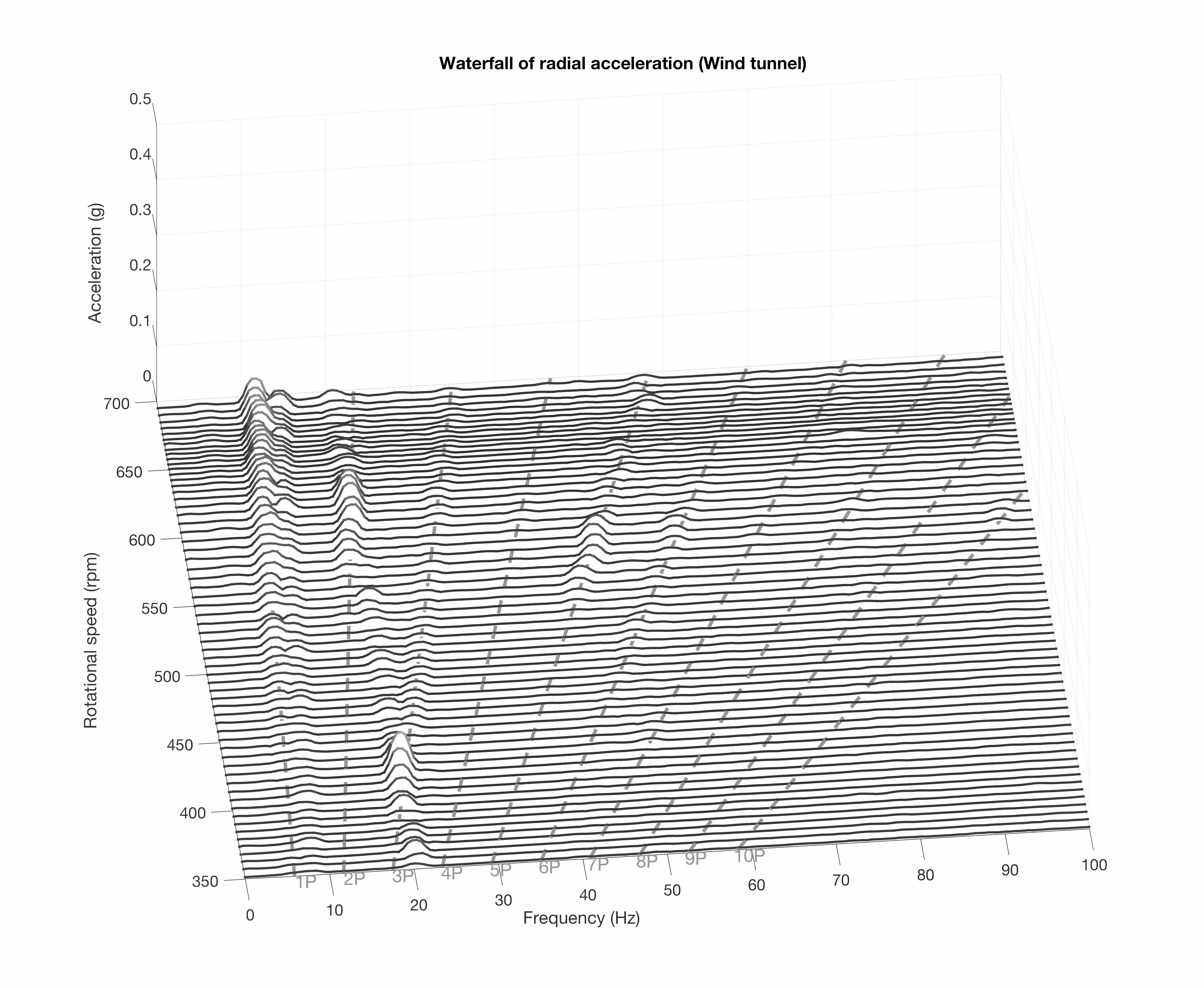}
    \caption{The waterfall plot of low frequency radial vibrations: undamaged generator.}
     \label{wat3}
  \end{center}
\end{figure}

In order to understand the main features of the vibration spectra, some results from the wind tunnel ramp tests are reported in the following waterfall plots. This kind of graphics is commonly used to represent data in the analysis of rotating machines in non stationary conditions when is necessary to study vibrations in many rotational speed regimes. In Y-axis are represented the rpm values in correspondence of which a spectral analysis is performed,as consequence in horizontal axis are represented the frequencies and in z-axis the amplitude of acceleration. In waterfall plots is then possible to visualize all the vibrating phenomena that are dependant by the rotational frequency or by one of its harmonic. According to this, in the following sections the shaft's frequency will be labeled as 1P, or first order, and the harmonics as 2p, 3P and so on. In the x-y plane of waterfall plots, orders are described by straight lines passing from the origin and with an angular coefficient of $60/n$, where $n$ is the number of the order.  In Figure \ref{wat1}, the waterfall plot of the vibrations under the ramp wind time series is reported, when the damaged generator is mounted in the HAWT. From Figure \ref{wat1}, a bird's eye view on the vibration spectrum can be obtained and the main observation is that the electromechanical coupling (48P and harmonics, as discussed in \cite{castellani2018experimental}) is the most important contribution.  Figures \ref{wat2} and \ref{wat3} are waterfall plots focused on the low-frequency components of the vibration spectrum, when respectively the damaged and the undamaged generator are mounted in the HAWT. Comparing Figure \ref{wat2} to Figure \ref{wat3}, it arises that the two frequency contents are different and this is a first meaningful information for the detection of the damage. 
 
In order to understand where the possible damage is located, it is necessary to elaborate more and to have a theoretical framework for connecting the frequency content to damages. The literature about bearing faults comes at hand \cite{qiao2015survey}. The bearing parameters are indicated as follows: ${ f }_{ r }$ is the number of rpm of the shaft (Hz), ${ D }_{ b }$ is the diameter of the rolling elements, ${ D }_{ p }$ is the mean diameter of the bearing (pitch diameter), $Z$ is the number of the rolling elements, $\alpha$ is the contact angle. The characteristic frequencies associated to possible faults are the following, in units of the fundamental rotational frequency (indicated in the following as 1P) or orders:

\begin{enumerate}
\item damage on the inner race (BPFI, Ball Pass Frequency on Inner ring): 
\begin{equation}
  f_{ i }=Z\frac {{ f }_{ r } }{ 2 } \left\{ 1+\frac { { D }_{ b } }{ { D }_{ p } } \cos { \left( \alpha  \right)  }  \right\}  
\end{equation}
\item damage on the outer race (BPFO, Ball Pass Frequency on Outer ring):
\begin{equation}
  f_{ o }=Z\frac { { f }_{ r } }{ 2 } \left\{ 1-\frac { { D }_{ b } }{ { D }_{ p } } \cos { \left( \alpha  \right)  }  \right\}  
\end{equation}
\item damage on the rolling elements (BSF, Ball Spin Frequency): 
\begin{equation}
 f_{ r }=\frac { { f }_{ r } }{ 2 } \frac { { D }_{ p } }{ { D }_{ b } } \left\{ 1-{ \left[ \frac { { D }_{ b } }{ { D }_{ p } } \cos { \left( \alpha  \right)  }  \right]  }^{ 2 } \right\}    
\end{equation}
\item damage on the cage (FTF, Foundamental Train Frequency): 
\begin{equation}
f_{ c }=\frac { { f }_{ r } }{ 2 } \left\{ 1-\frac { { D }_{ b } }{ { D }_{ p } } \cos { \left( \alpha  \right)  }  \right\}    
\end{equation}
\end{enumerate}

For the considered test case, $\alpha=0$, $Z=9$, $D_b=0.0111$ meters, $D_p=0.0549$ meters and the following orders are obtained:

\begin{table}[H]
\centering
\caption{Characteristic order frequencies for bearing faults.}
\begin{tabular}{c|cc}
damaged component   & order  (-)  \\
\hline
inner race &    5.41        \\
outer race  &    3.59       \\
rolling elements  &    2.37       \\
cage  &    0.40     \\
\hline
\end{tabular}
\label{freq}
\end{table}

\begin{figure}[H]
\begin{center}
\includegraphics[width=120mm]{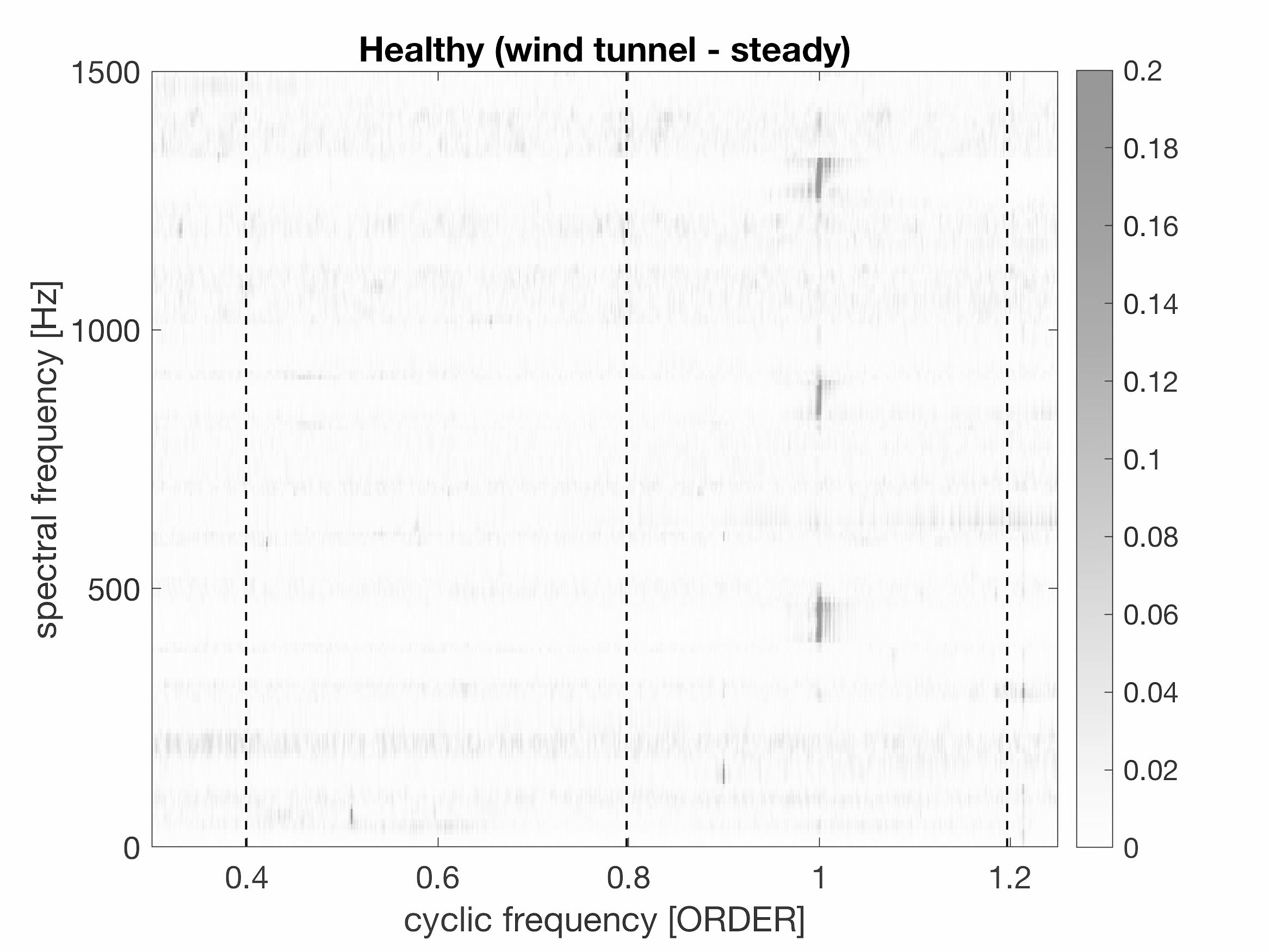}
\caption{Spectral correlation function of wind tunnel vibration measurements at 400 rpm: undamaged generator.}
\label{spec1}
\end{center}
\end{figure}

\begin{figure}[H]
\begin{center}
\includegraphics[width=120mm]{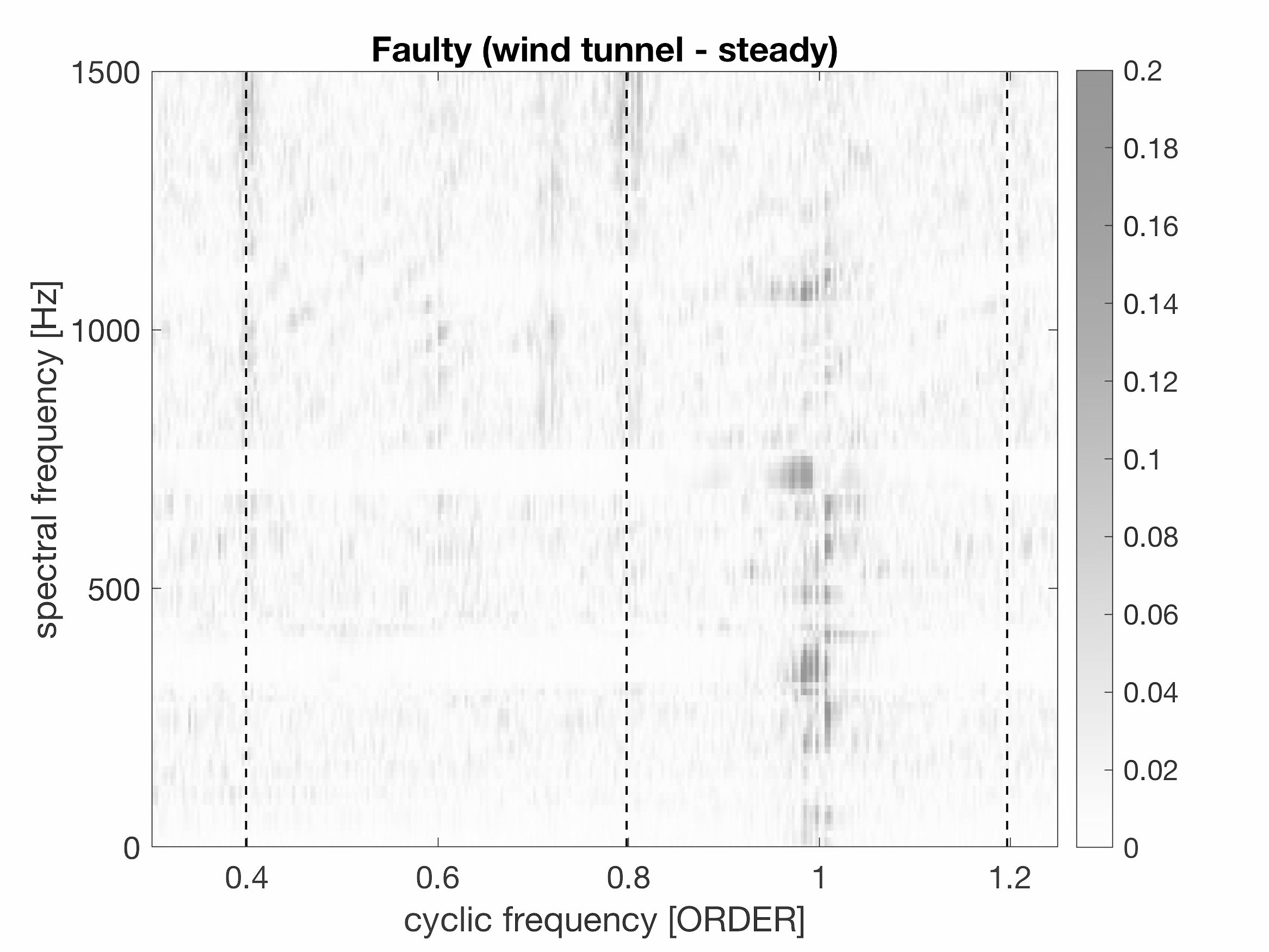}
\caption{Spectral correlation function of wind tunnel vibration measurements at 400 rpm: damaged generator.}
\label{spec2}
\end{center}
\end{figure}

In the frequency domain, cyclostationarity \cite{antoni2007cyclic} is a very powerful approach for characterizing non-stationary signals of rotating machinery as wind turbines.  The analysis of the spectral correlation function is particularly adequate for interpreting the frequency content of a signal:
\begin{equation}
S_{XX}\left(\alpha, f\right)=\lim_{T\to\infty} \EX \left\{X_T\left(f+\frac{\alpha}{2}\right) X^*_T\left(f-\frac{\alpha}{2}\right)\right\},
\end{equation}
where $X_T(f)$ is the Fourier transform of the signal $x(t)$ over an interval $T$. If the modulation of the signal is periodic, the spectral correlation is continuous in $f$ and discrete in $\alpha$: it is non-zero only at those frequencies that are multiple of the fundamental modulation frequency (the 1P, in this case). Therefore, if the generator has a damage at the bearing, the spectral correlation function should be non-zero for some $\alpha$ indicated in Table \ref{freq} (and higher harmonics).  This results indeed to be the case, as can be seen in Figures \ref{spec1} and Figure \ref{spec2}, where the x-axis is the order of the frequency.

Comparing the two figures, it arises that in Figure \ref{spec2} a contribution at 0.4P (and second harmonic) is visible: it corresponds to the characteristic frequency of the cage (see Table \ref{freq}). This points to a possible detection of the damage. Further support to this hypothesis comes from a similar analysis of the test rig data.

\begin{figure}[H]
\begin{center}
\includegraphics[width=120mm]{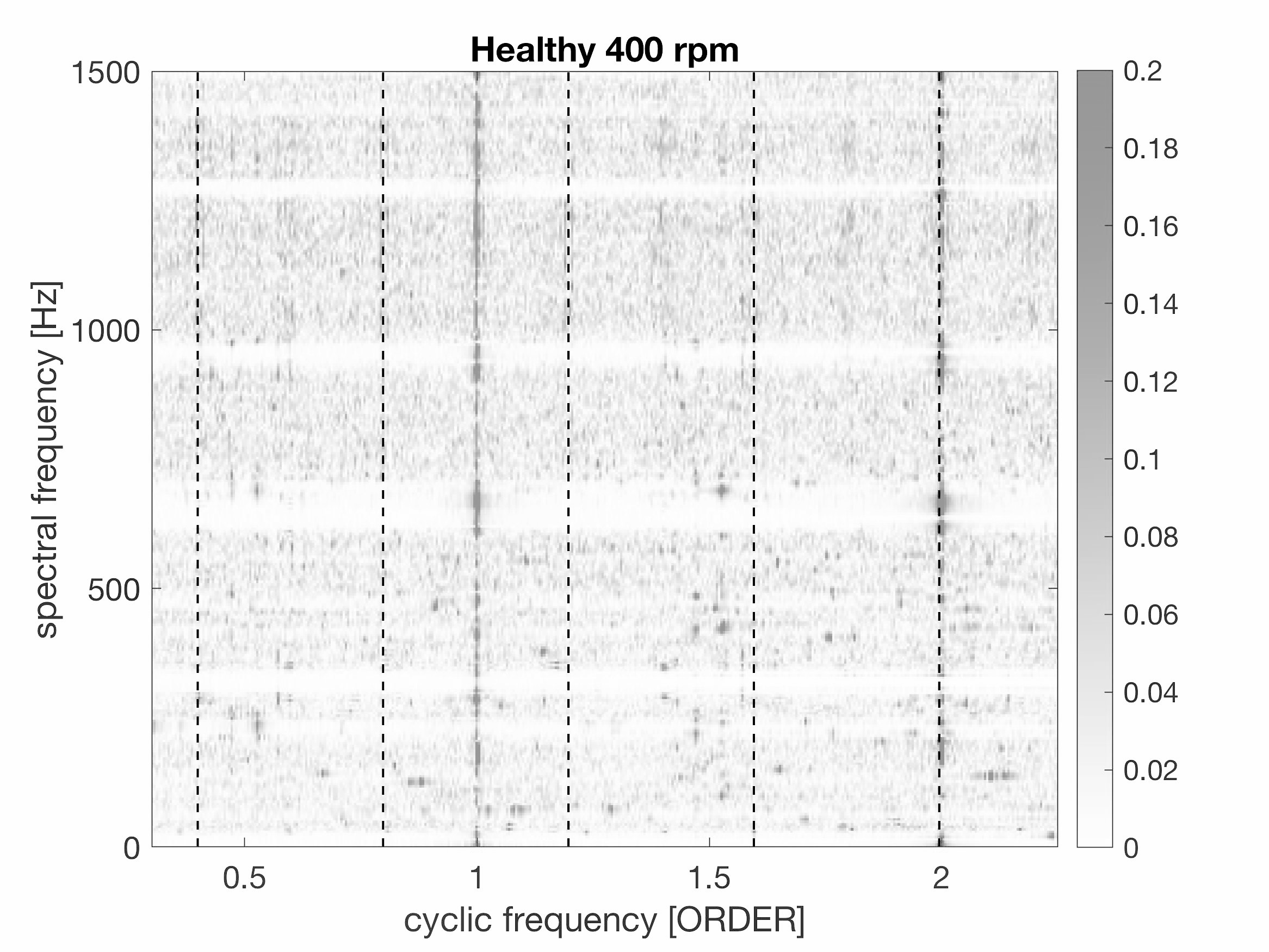}
\caption{Spectral correlation function of test rig vibration measurements at 400 rpm: undamaged generator.}
\label{spec3}
\end{center}
\end{figure}

\begin{figure}[H]
\begin{center}
\includegraphics[width=120mm]{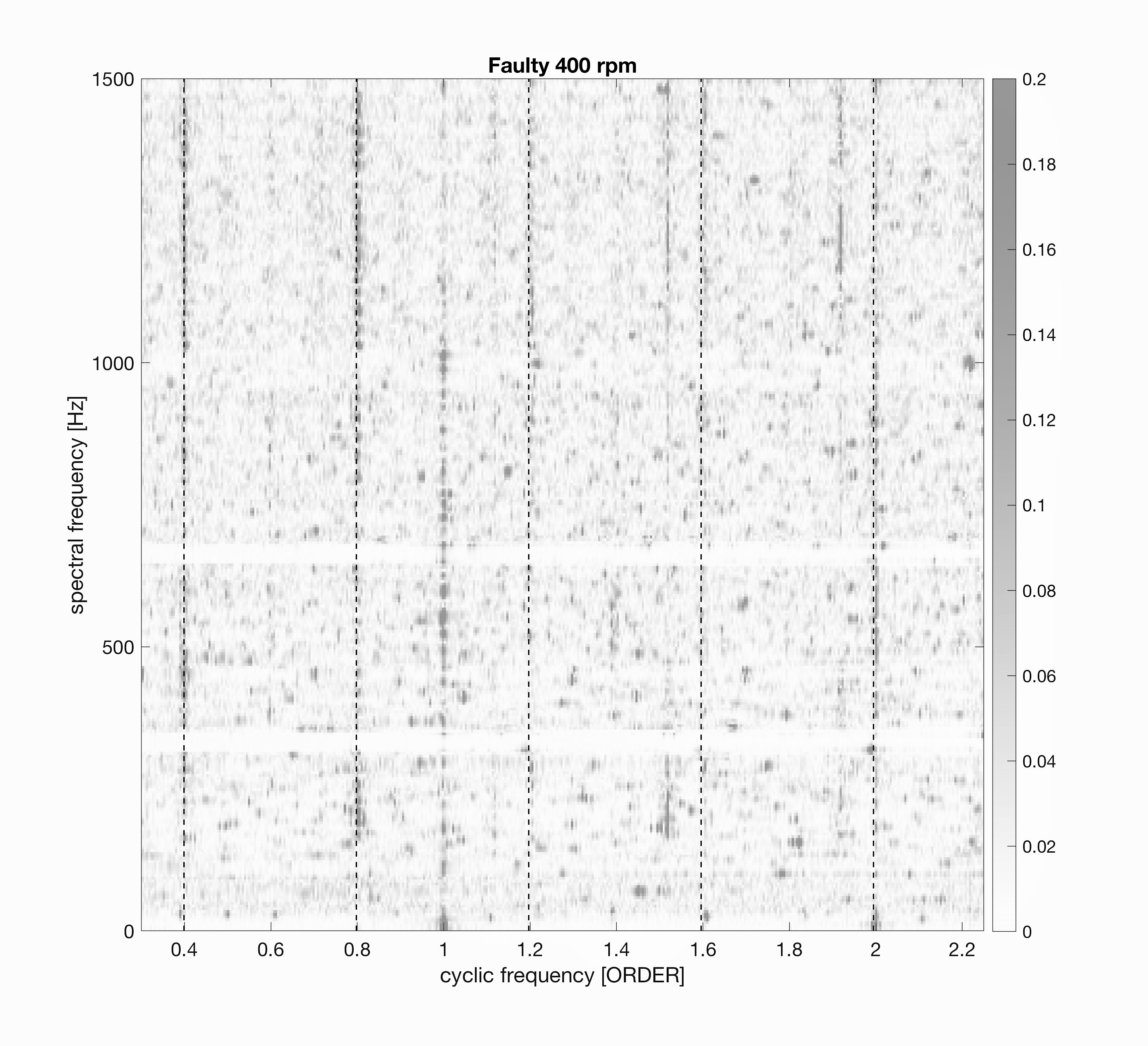}
\caption{Spectral correlation function of test rig vibration measurements at 400 rpm: damaged generator.}
\label{spec4}
\end{center}
\end{figure}

The spectral correlation function for the test rig measurements at 400 rpm is reported in Figures \ref{spec3} and Figure \ref{spec4} and it confirms the indications arising from Figures \ref{spec1} and \ref{spec2}: the 0.4P order contributions is visible in the tests of the damaged generator and this is compatible with a defect at the bearing cage. In addition, the frequencies multiples of 0.4P, especially 0.8P, are clearly visible.

The test rig data have been employed for a further analysis in the time domain. The squared Mahalanobis distance is selected as an appropriate metric for signal feature extraction \cite{daszykowski2007robust, wu2013multi, de2018wind}: it is a multidimensional generalization of the Euclidean distance, that measures the number of standard deviations between a given point and a distribution. Furthermore, it takes into account the correlation among the data sets. Given a set of reference measurements $\left(x_1, \dots, x_n\right)$, whose average is $\bar{x}$ and whose covariance matrix is $S$, the Mahalanobis distance between one measure $y$ and the $x$ distribution is given by
\begin{equation}
\label{maha}
d_M(y)=\sqrt{\left(y-\bar{x}\right) S^{-1} \left(y-\bar{x}\right)}
\end{equation}
The Mahalanobis distance is employed as follows. Tests at 400 rpm are selected and one time series from the undamaged generator is taken as reference (i.e. as $x$ in Equation \ref{maha}). The target time series are collected as well at 400 rpm and one of them is taken from the undamaged generator and one is taken from the damaged one. For each point of the two target time series, the Mahalanobis distance with respect to the reference $x$ set is computed and therefore two vectors of Mahalanobis distances are obtained. 

\begin{figure}[H]
\begin{center}
\includegraphics[width=100mm]{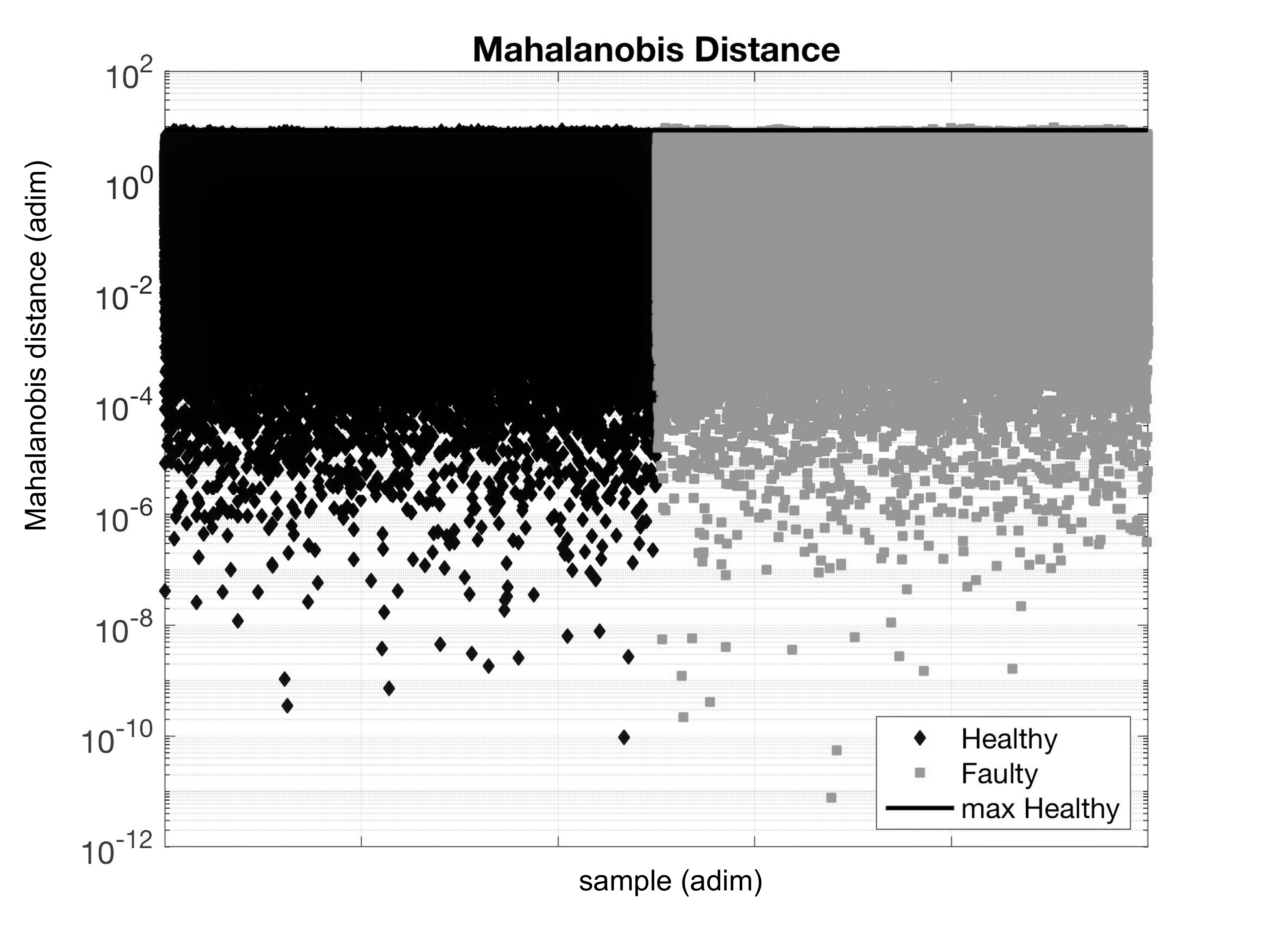}
\caption{Mahalanobis distance between reference time series of undamaged generator and target time series of undamaged (black) and damaged (grey) generator. Test rig measurements at 400 rpm.}
\label{mahal1}
\end{center}
\end{figure}

In Figure \ref{mahal1}, results are reported for the study of non-processed data in the time domain: no clear indications arise for distinguishing the damaged from the undamaged generator.

\begin{figure}[H]
\begin{center}
\includegraphics[width=100mm]{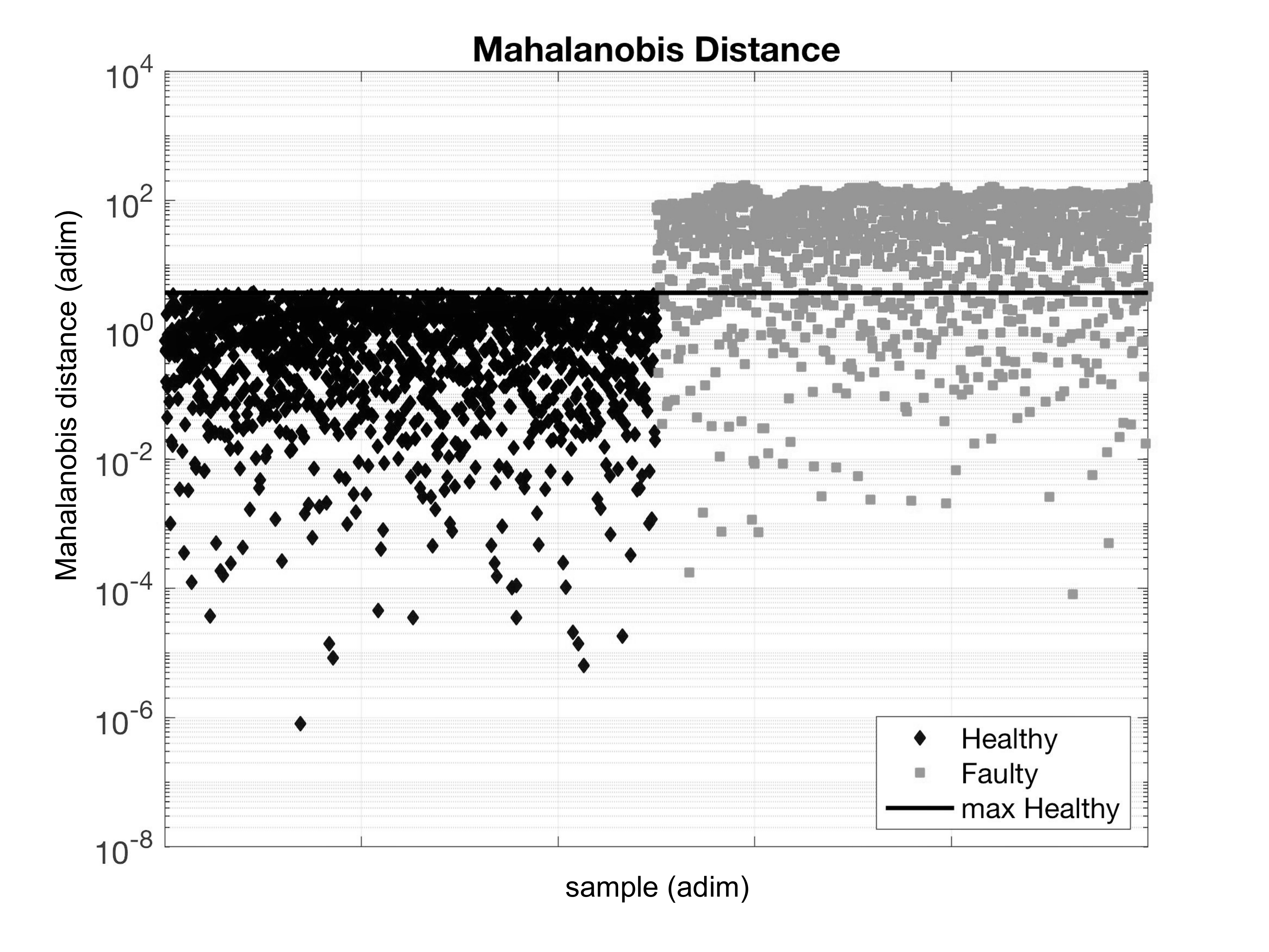}
\caption{Mahalanobis distance between reference time series of undamaged generator and target time series of undamaged (black) and damaged (grey) generator. Data are order-tracked around the cage characteristic frequency. Test rig measurements at 400 rpm.}
\label{mahal2}
\end{center}
\end{figure}

\begin{figure}[H]
\begin{center}
\includegraphics[width=100mm]{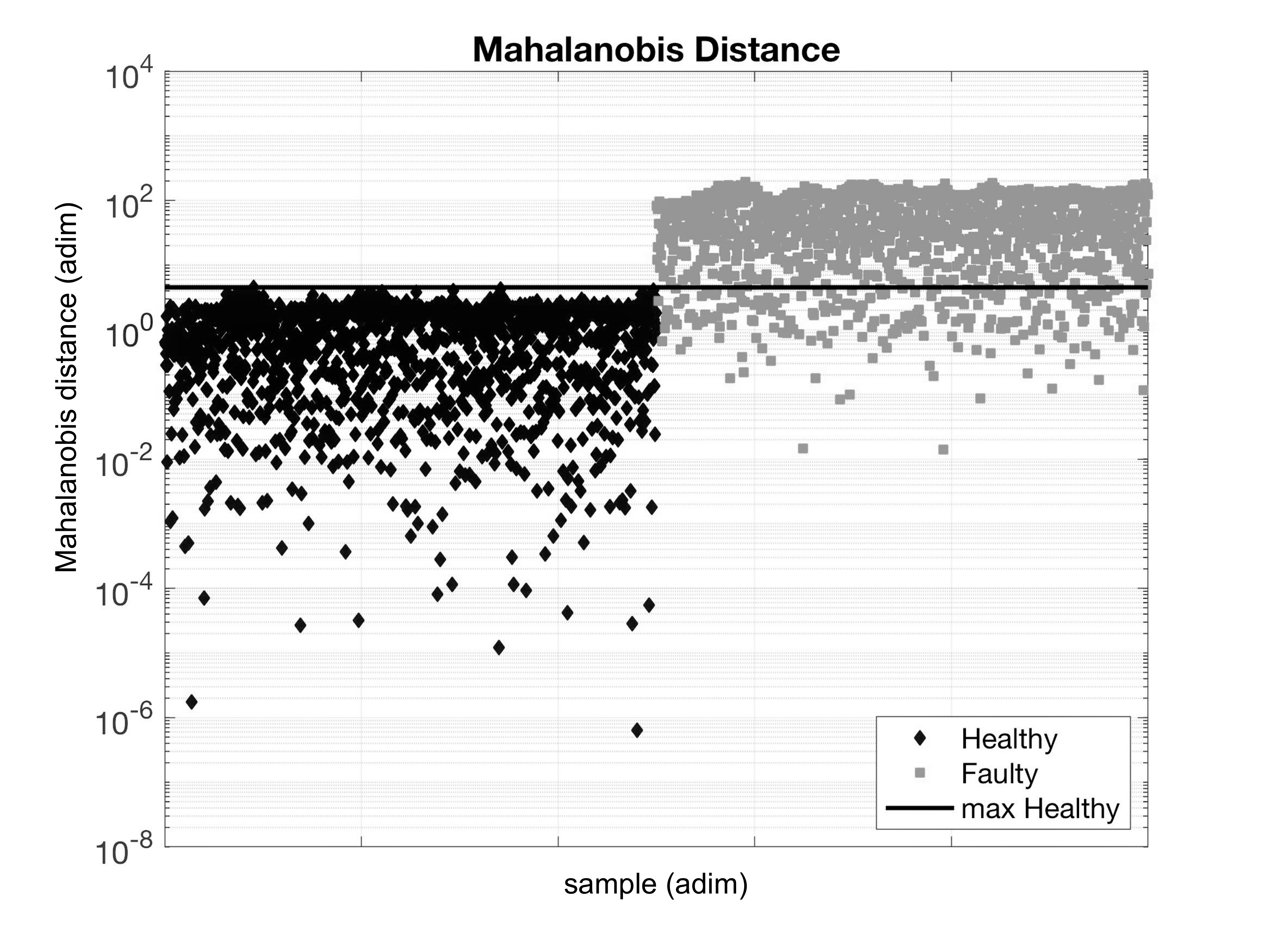}
\caption{Mahalanobis distance between reference time series of undamaged generator and target time series of undamaged (black) and damaged (grey) generator. Data are order-tracked around the inner race characteristic frequency. Test rig measurements at 400 rpm.}
\label{mahal3}
\end{center}
\end{figure}

Since Figure \ref{mahal1} results to be indecisive, data have subsequently order tracked around the characteristic frequencies of the bearing (Table \ref{freq}) and sample results are reported in Figures \ref{mahal2} and \ref{mahal3}, respectively for the cage and inner track frequency. 

From Figures \ref{mahal2} and \ref{mahal3}, it is possible to clearly distinguish the data of the damaged generator with respect to the undamaged one. Since the data have been order-tracked around two characteristic frequencies of the bearing for obtaining Figures \ref{mahal2} and \ref{mahal3}, this kind of analysis confirms the fact that the damage regards the generator bearing. The weak point is that Figures \ref{mahal2} and \ref{mahal3} are basically indistinguishable, but they have been obtained filtering the data around two different bearing defect orders (cage and inner race). Therefore, this kind of analysis seems to be weaker than ciclostationarity for identifying the precise location of the damage in the bearing.

\section{Discussion}
\label{disc}

In the present study, the use of several experimental and post-processing techniques for condition monitoring of small HAWT generators has been explored. The test case of interest is a 1.8 kW PMGs: tests have been performed using a damaged and an undamaged one.

The two selected experimental techniques are wind tunnel and test rig measurements. The difference between the two is, evidently, that wind tunnel tests involve the entire device (small HAWT) in operation, while the test rig measurement campaign is focused on the sub-component of interest. Therefore, the main issue faced in this work is the following: can it be possible to do condition monitoring of small HAWT through wind tunnel tests? In other words, has it been possible, for the test case of this work, to extract from the wind tunnel measurements and the test rig measurements the same information about the damage location? The answer is positive, but it depends on the selected post-processing techniques. 

The reason is that the wind tunnel measurements involve the whole HAWT device operating and interacting with the wind: contributions to the spectrum arise from the periodic motion of the blades, can arise from rotor imbalance effects (also causing magnetic pull on the generator and therefore amplifying the electromechanical coupling), and so on. All these effects can possibly mix in the spectrum with the contributions from the damage: as can be seen in Table \ref{freq}, the orders of bearing damages can be relatively close to 3P or 6P. The downside of wind tunnel measurements is therefore that it is difficult to use this kind of data for a decisive localization of a damage; the advantage is practical (the device can be tested in its entirety) and furthermore this kind of tests provide a bird's eye view on the dynamic behavior of the system, that can be useful for the damage diagnosis. The test rig data, instead, are more adequate for investigating definite portions of the vibration spectrum because the subcomponent is isolated from the rest of the device: therefore, the final diagnosis of a mechanical damage in a HAWT is more decisively obtained using test rig data, but adequate post-processing techniques of the wind tunnel measurements provide very useful information as well.

The results in Section \ref{S3} indicate that the feature extraction in the time domain can be prohibitive even when processing the test rig measurements (Figures \ref{mahal1}, \ref{mahal2}, \ref{mahal3}). A time-frequency approach is needed and cyclostationarity results being the optimal framework. Actually the low-frequency tail of the waterfall plot (Figures \ref{wat2} and \ref{wat3}) of the wind tunnel measurements presents a clear distinction between the vibration scenarios arising when the damaged and the undamaged generators are adopted, but it doesn't help for identifying the damage location. Instead, the spectral correlation function plot provides meaningful indications about the damage location and this happens for the wind tunnel (Figures \ref{spec1} and  \ref{spec2}) and test rig measurements (Figures \ref{spec3} and \ref{spec4}): the distinction between the two is that the test rig data are clearer and easier to interpret.

\section{Conclusions}
\label{S4}

This work was devoted to the experimental analysis of a permanent magnet generator of a small HAWT, with the aim of practicing condition monitoring and damage diagnosis techniques. It is remarkable that this topic is quite overlooked in the scientific literature \cite{cai2016condition}, despite the fact that the peculiarity of small HAWTs (small dimensions and high rotational speed) call for sophisticated analysis techniques. This kind of studies have a technological impact as well, because the exploitation of small HAWTs, especially in urban environment, poses several challenges in the design and prototyping phase, as regards control and noise and vibration mitigation.

The test case of this work was a 1.8 kW PMG, that has been adopted on a HAWT having 2 meters of rotor diameter and 3 kW of maximum producible power. Two identical models of the PMG have been considered: one is suspected to be damaged and one is undamaged.  Experimental data have been collected through accelerometers in the wind tunnel, under steady and ramp tests, and on a test rig driven at several shaft rotational speeds. The test rig measurements resulted to be useful in the early diagnosis phase, because they are particularly adequate for quantifying the performance loss of the damaged generator with respect to the undamaged one.

Since the generator is gearless, the analysis has been focused on the bearings and the characteristic frequencies associated to damages on the different parts have been identified. The test rig data have been particularly useful also for the final diagnosis of the damage, because the spectrum can be easier interpreted with respect to what happens with wind tunnel test involving the whole HAWT device in motion.

Furthermore, an important outcome of this test case is that, despite the analysis in the time domain provides some indications of the anomaly, the damage can be better localized using signal analysis techniques devoted to cyclostationary phenomena. It is remarkable that, studying the behavior of the spectral correlation function, it has been possible to identify a frequency content compatible with a damage on the cage of the bearing, using the test rig data as well as the wind tunnel data (involving the whole device).
Mahalanobis distance turned to be a useful instrument to detect faults in the generator even if data has to be order-tracked around characteristic frequencies.    

The general lesson from this study is that small HAWT technology can be troublesome as regards performances and noise and vibration issues. The wind tunnel measurements on this kind of devices have the great pro that they provide a bird's eye view on the dynamic behavior of the system. Their use is therefore useful, but it might be not sufficient for the precise localization of a damage or for precision condition monitoring. The test rig measurements, instead, are more adequate for investigating definite portions of the vibration spectrum: therefore, the final diagnosis of a mechanical damage in a small HAWT is likely to be obtained using test rig data. Small HAWT condition monitoring therefore requires complex techniques and experimental facilities, whose study should be encouraged in the academia and whose adoption should be boosted in the industry for the successful use of this kind of clean energy conversion systems. 

One important further direction of this work is improving the post-processing techniques in order to extract more information from the measurement campaigns: actually, with the present study it has been possible to detect and reasonably locate the bearing damage, but it hasn't been possible to estimate the remaining useful life of the sub-component. This information is indeed important when dealing with operation and maintenance of energy conversion devices. Therefore, it desirable to develop post-processing techniques similar, for example, to those in \cite{cao2018prediction}. Another interesting further direction of this work is performing sound and vibration measurement campaigns: if successful, this would allow exploring the possibility of elaborating information about the health state of the small HAWT using a relatively simple experimental set up.

\vspace{6pt} 





\acknowledgments{This research activity was partially supported by Italian PRIN funding source (Research Projects of National Interest - Progetti di Ricerca di Interesse Nazionale) through a financed project entitled SOFTWIND (Smart Optimized Fault Tolerant WIND turbines).}


\abbreviations{The following abbreviations are used in this manuscript:\\

\noindent 
\begin{tabular}{@{}ll}
CM & Condition Monitoring\\
HAWT & Horizontal-Axis Wind Turbine\\
PMG  & Permanent Magnet Generator \\
rpm & Revolutions Per Minute
\end{tabular}}




\reftitle{References}


\externalbibliography{yes}
\bibliography{biblio}



\end{document}